\documentclass{jfm}

\usepackage{graphicx}
\usepackage{amssymb}
\usepackage{amssymb}
\usepackage{newtxtext}
\usepackage{newtxmath}
\usepackage{natbib}
\usepackage{hyperref}
\hypersetup{
    colorlinks = true,
    urlcolor   = blue,
    citecolor  = black,
}

\newcommand{\RomanNumeralCaps}[1]
\linenumbers

\title{Resolvent analysis of a swimming foil}

\author{J. M. O. Massey\aff{1,2,3}
  \corresp{\email{masseyj@stanford.edu}},
  S. Symon \aff{1}
  B. Ganapathisubramani\aff{1}
  \and G. D. Weymouth\aff{1,2}}

\affiliation{\aff{1}Faculty of Engineering and Physical Sciences, University of Southampton, UK
\aff{2}Faculty of Mechanical Engineering, TU Delft, NL
\aff{3}Department of Mechanical Engineering, Stanford University, Stanford, CA 94305, USA}
\begin{document}
\maketitle

\begin{abstract}

    This study employs resolvent analysis to explore the dynamics and coherent structures in the boundary layer of a foil that swims via a travelling wave undulation. A modified NACA foil shape is used together with undulatory kinematics to represent fish-like bodies at realistic Reynolds numbers ($ \Rey = 10,000 $ and $ \Rey = 100,000 $) in both thrust- and drag-producing propulsion regimes.
    We introduce a novel coordinate transformation that enables the implementation of the data-driven resolvent analysis \citep{herrmann_data-driven_2021} to dissect the stability of the boundary layer of the swimming foil.
    This is the first study to implement resolvent analysis on deforming bodies with non-zero thickness and at realistic swimming Reynolds numbers.
    The analysis reinforces the notion that swimming kinematics drive the system's physics. In drag-producing regimes, it reveals breakdown mechanisms of the propulsive wave, while thrust-producing regimes show a uniform wave amplification across the foil's back half. The key thrust and drag mechanisms scale with the boundary-layer thickness, implying geometric self-similarity in this $\Rey$ regime. In addition, we identify a mechanism that is less strongly coupled to the body motion.
    We offer a comparison to a rough foil that reduces the amplification of this mechanism, demonstrating the potential of roughness to control the amplification of key mechanisms in the flow.
    The results provide valuable insights into the dynamics of swimming bodies and highlight avenues for developing opposition control strategies.
    
\end{abstract}


\section{Introduction}

    The study of the boundary layer of swimming bodies has been addressed in the literature by \citet{anderson_boundary_2001}, who investigated the boundary layer of a waving plate under various swimming conditions. They were unable to provide information on the dynamics or coherent structures in the boundary layer, information crucial to understanding the system \citep{rempfer_evolution_1994}.
    
    Swimming involves creating a propulsive wave through a time-periodic body deformation.
    Resolvent analysis \citep{mckeon_critical-layer_2010} has had vast success in uncovering driving mechanisms in complex flows. The analysis treats the non-linearity in the fluctuating part of the Navier-Stokes equation as an unknown harmonic forcing. 
    However, the traditional formulation linearises around the time-averaged mean flow to study the transfer of energy from the mean to dominant fluctuations. As such, only the first harmonic is identified in oscillator-type flows, whereas higher harmonics are not, since they receive energy from the first harmonic and not the mean \citep{symon_tale_2019}. To improve predictions at higher harmonics, the linearisation can be performed around an alternative base flow. For example, \citealt{padovan_analysis_2020} linearised around a Fourier basis to offer insights into the unstable dynamics of the flow over a NACA0012 aerofoil at $\Rey=200$ and an angle of attack of $20^{\circ}$.

    \citet{herrmann_data-driven_2021} adopts a data-driven approach to obtain the resolvent operator, that uses the link between Dynamic Mode Decomposition (DMD) and Koopman eigenfunctions \citep{williams_datadriven_2015} to construct the linear basis for analysis. DMD, introduced by \citet{schmid_dynamic_2010}, enables the extraction of spatio-temporal patterns from time-resolved data. The challenge of a moving and deformable body is that the domain position containing the fluid and the body depends on time. \citet{goza_modal_2018} successfully coupled the fluid and body domains to identify the driving mechanisms for the onset of chaotic flapping, however, their body was an infinitesimally thin membrane, so they did not have to consider the temporal domain switch between the fluid and body regions. \citet{menon_dynamic_2020} looked at the pitching and plunging of an aerofoil by rotating and translating the reference frame, so the body remained centred within the reference frame. This is limited in that it deals with a rigid body.

    In this paper, we investigate the dominant mechanisms in the boundary layer of a swimming foil using the data-driven resolvent analysis \citep{herrmann_data-driven_2021}. We use a modified NACA foil shape to represent a fish-like body and a generalised swimming motion to represent the kinematics. The foil swims at both $\Rey=10,000$ and $\Rey=100,000$. We consider thrust- and drag-producing propulsive regimes. We propose a novel coordinate transform to study the boundary layer of a swimming foil in body coordinates.
    Finally, we provide a comparison to a rough foil that reduces the amplification of a mechanism at the third harmonic of the swimming frequency.
    To our knowledge, this is the first study of its kind to implement resolvent analysis on deforming bodies with nonzero thickness. In addition, we uncover key physical differences between thrust- and drag-producing foils and highlight coherent structures that scale with the boundary-layer thickness.

\section{Methodology}
    \subsection{Geometry and kinematics}\label{sec:RA geometry}
        \begin{figure}
            \centering
            \includegraphics[width=\textwidth]{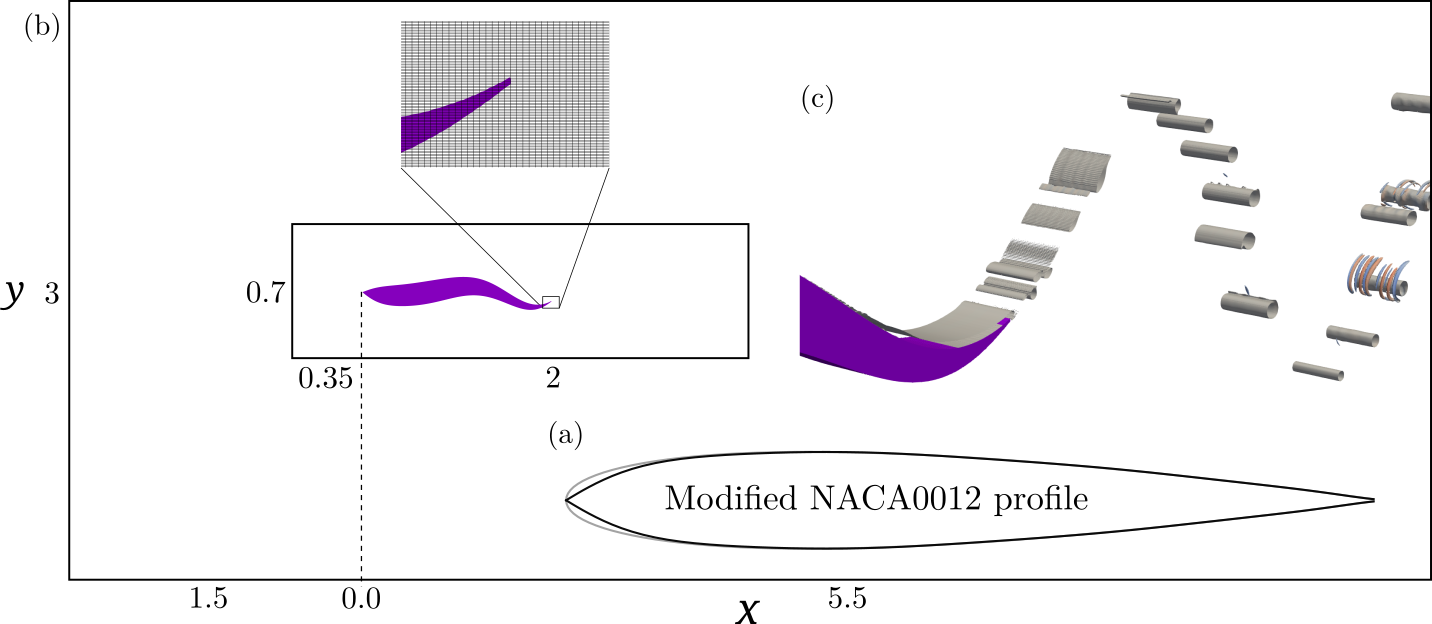}
            \caption{Illustration of the computational set up and validation. (a) Comparison of the NACA0012 profile in grey and the modified profile plotted on top (black). (b) Schematic of the domain and grid. The inner box shows the region where the grid is uniformly rectilinear; from there the grid stretches toward the domain extent. The inset shows the grid resolution around the tail of the foil. (c) Vortex structures in the wake visualised by the $1\times 10^{-5}$ contour of the Q-criterion, coloured by the y-vorticity.}
            \label{fig:domain}
        \end{figure}

        We use a NACA0012 profile, modified with a sharpened leading edge, to represent a fish-like body. Figure \ref{fig:domain} compares the modified profile with the NACA0012 excursion data.

        Scaling all lengths by $L$, using the swimming speed $U$ to scale the velocity and $L/U$ to scale the time, we define the excursion of the body from the centre line as        
        	\begin{equation}
        	    y(x) = A(x) \sin \big(2\pi  \, [f t - x/\zeta] \big),
            \label{eq:swimming}
        	\end{equation}
        where $f$ is the frequency, $\zeta$ is the phase speed of the travelling wave, and $A(x)$ is the amplitude envelope. The Strouhal number is set to a value of $St=0.3$, which determines the scaled frequency as $f=St/2A_1$, where $A_1=A(x=1)$ is the trailing edge amplitude and defines the wake width of the system. We use the result from \citet{di_santo_convergence_2021} for the envelope
        \begin{equation}\label{eq:kinematic trajectory}
            A(x)=\frac{A_1(a_2x^2+a_1x+a_0)}{\sum_{i=0}^{2} a_i},
        \end{equation}
        using $a_{0,1,2}=(0.05, -0.13, 0.28)$. $A_1=0.1$, which was found to be optimal in \citet{Saadat2017}. Finally, we set the wave speed to place the foil in a thrust or drag-producing regime (figure \ref{fig:verf} a). For the thrust-producing foil $\zeta=1.25$ and for the drag $\zeta=1.75$. For the convergence analysis, we use a representative simulation using $\zeta=1.42$, which we find to result in a cycle averaged zero net thrust.

    \subsection{Numerical method}\label{sec:numerical method}
    
        We simulate incompressible fluid flow with the dimensionless Navier-Stokes combined with the continuity equation        
        \begin{align}
            \frac{\partial \mathbf{u}}{\partial t} + (\mathbf{u} \cdot \mathbf{\nabla})\mathbf{u} = &
            -\mathbf{\nabla}p + \frac 1\Rey \nabla^2 \mathbf{u}, \\
            \mathbf{\nabla} \cdot \mathbf{u} = & 0 \text{ ,}
        \end{align}
        where $\mathbf{u}(\mathbf{x}, t) = (u, v, w)$ is the scaled velocity of the flow and $p(\mathbf{x},t)$ is the scaled pressure. We use an in-house developed solver based on the Boundary Data Immersion Method (BDIM) \citep{maertens_accurate_2015}. The solver has been extensively validated in swimming studies \citep{maertens_optimal_2017, Zurman-Nasution2020, massey_systematic_2023, lauber_rapid_2023} and converges at second order in both time and space \citep{lauber_immersed_2022}.
        
        The boundary conditions we enforce on the body is the no-slip condition. Symmetry conditions are enforced on the upper and lower domain extents, and there is a periodic condition in the spanwise direction. Figure \ref{fig:domain} details the grid and domain configuration. We use the smallest domain for which the forces on the body remain invariant when the size increases. We use the thrust force based on the surface pressure, $C_T = \oint \mathbf{f_x} ds/0.5 A_1$, where $\mathbf{f}=-p\hat n$ is the normal pressure stress on the body.

        \begin{figure}
            \centering
            \includegraphics{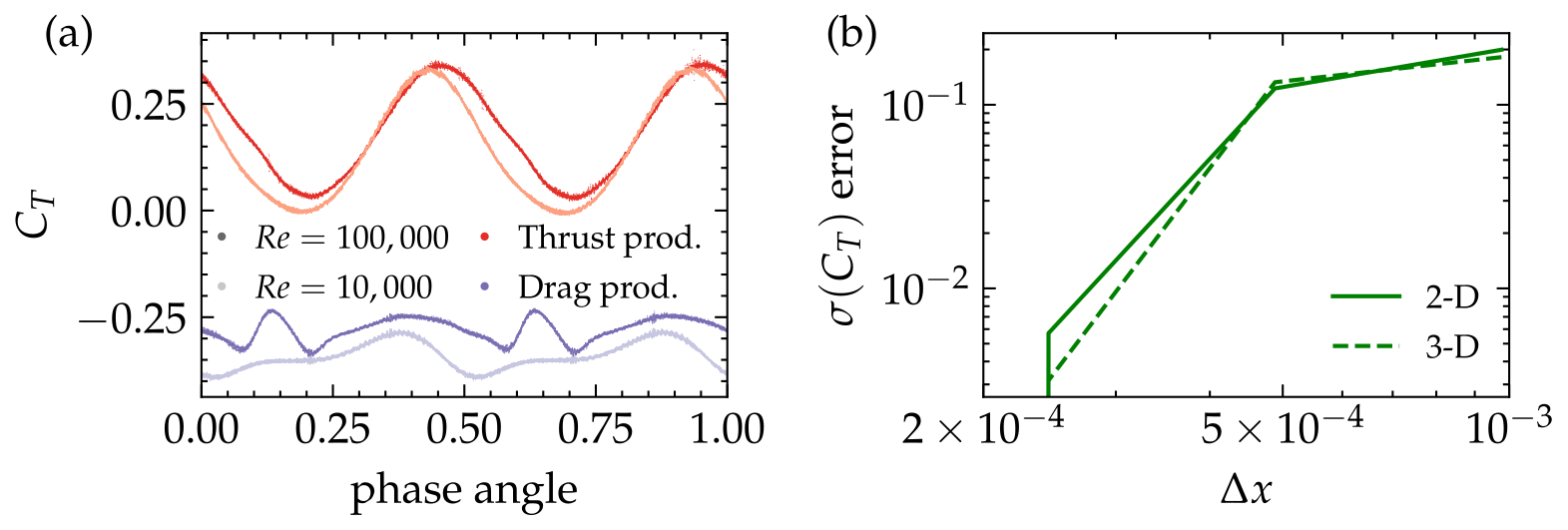}
            \caption{(a) Cycle normalised thrust coefficients for the 4 test cases. The red lines are the thrust producing cases and the purple are the drag producing cases. $\Rey$s are differentiated by the intensity of the colour, as labelled. (b) Convergence of the standard deviation of the thrust coefficient for a representative simulation with zero net thrust. The error of both 2-D and 3-D simulations are relative to a 2-D simulation with $\Delta y=1/8192$.}
            \label{fig:verf}
        \end{figure}
        
        We use a rectilinear grid in the domain $x\in[-1.5, 5.5]$, $y\in [-1.5,1.5]$, and $z\in [0,1/64]$ (figure \ref{fig:domain}a). For the area that contains the body motion and the immediate wake, we use a uniform grid and then implement hyperbolic stretching of the grid cells away from this area (figure \ref{fig:domain}). The grid is refined in $y$ such that $\Delta y$ is a quarter of $\Delta x$ and $\Delta z$ (figure \ref{fig:domain} inset). The total number of grid cells is $ N=(4096, 4096, 64)$ totalling $N=1.07\times10^9$ grid cells.
        
        The grid resolution is shown to be sufficient by using representative two-dimensional and three-dimensional simulations. Using 2-D simulations allows us to increase the simulation comparison point one extra refinement step and check the resolution at the final working resolution is sufficient. Figure \ref{fig:domain}c shows that the flow has little variation in the spanwise direction similar to the results shown in \citet{Zurman-Nasution2020, massey_systematic_2023}. Therefore, we use a highly refined two-dimensional simulation as a final comparison point for both the 2-D and 3-D cases, avoiding the need for the extremely large fine grid in 3-D. Figure \ref{fig:verf} b shows that the standard deviation of the thrust coefficient converges for both 2-D and 3-D. The domain size is validated by comparing the thrust with the phase angle for the working domain and a much larger domain with dimensions $x\in[-5,20]$ and $y\in [-5,5]$; the thrust is invariant with the phase angle (not shown), indicating that the domain size is sufficient. Finally, we save 800 snapshots of the velocity field over 4 cycles for the subsequent analysis.

    \subsection{Coordinate transform}\label{sec:coordinate transform}
    
        \begin{figure}
            \centering
            \includegraphics{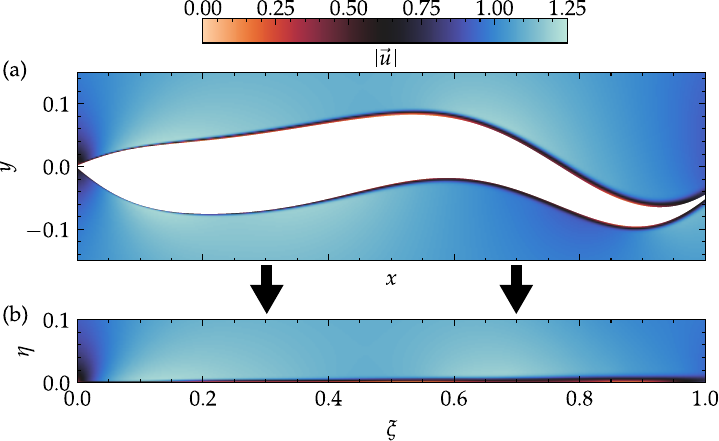}
            \caption{The velocity magnitude in (a) Cartesian coordinates from the computational domain and (b) the local normal-tangential coordinates on the upper section of the foil.}
            \label{fig:transform}
        \end{figure}

        DMD produces inaccurate modes around a body with motion \citep{menon_dynamic_2020}. This is because the temporal switch between the body and the fluid domain causes the DMD to scatter the coherent structures in the boundary layer across the spatial domain. The analysis can still be performed on the wake but misses valuable insights into the boundary-layer dynamics. Instead, we transform the flow into local normal-tangential coordinates to investigate the stability of the boundary layer of the swimming foil. This enables us to identify spatially and temporally coherent structures through DMD.

        Addressing the limitations of Dynamic Mode Decomposition (DMD) around moving bodies, we employ a transformation to local normal-tangential coordinates, as illustrated in figure \ref{fig:transform}. For simplicity, we spanwise average the flow. This approach results in a loss of information about spanwise mechanisms; however, we consider this loss to be minimal due to the small spanwise variability on the foil (figure \ref{fig:domain}b), and the fact that spanwise fluctuations occur predominantly in the wake of the foil. The global Cartesian coordinates are transformed into a local normal-tangential coordinate system through a mapping $\mathcal{T}$ such that
        \begin{equation}
            \mathcal{T} : \mathbf{x} \rightarrow \mathbf{\xi} \text{ ,}
        \end{equation}
        \
        where the Cartesian and local normal-tangential coordinates are        
        \begin{equation}
            \mathbf{x}=\begin{pmatrix} x \\ y \end{pmatrix},  \quad \mathbf{\xi}=\begin{pmatrix} \xi \\ \eta \end{pmatrix} \text{ ,}
        \end{equation}
        with $\xi$ aligned tangentially and $\eta$ aligned normally to the foil surface. In this analysis, we will only consider the top boundary layer of the foil, so the transformation can be determined by considering a superposition between the polynomial describing the shape of the foil and the equation \ref{eq:swimming} describing the motion. The position of the body boundary is then defined by $\mathbf{d}(t)$. The unit normal vector ($\hat{n} = \nabla \mathbf{d}/\left\| \nabla \mathbf{d} \right\|$) is calculated with respect to spatial coordinates, establishing a normal direction at each point along the body's surface, thus facilitating the mapping ($\mathcal{T}(\mathbf{d}, \eta)$) between Cartesian and local normal-tangential frames        
        \begin{equation}
            \mathcal{T} = \mathbf{d} + \hat{n}\eta \text{ ,}
        \end{equation}
        which satisfies $\xi=x|_{\eta=0}$ on the body. The normal component ($\eta$) is uniformly sampled, simplifying the interpolation process without significantly impacting the accuracy of the analysis. The velocity field in the local normal-tangential coordinates ($\mathbf{u}_\mathcal{B}$) is obtained by an interpolation of the velocity in the Cartesian coordinates ($\mathbf{u}$) at the local normal-tangential sample points
        \begin{equation}
            \mathbf{u}_\mathcal{B} = \mathcal{I}\left(\mathbf{u}_{\text{nn}}, (\eta, \xi)\right) \text{ ,}
        \end{equation}
        where $\mathcal{I}$, a linear interpolation operator chosen for its computational efficiency and accuracy, maps the nearest-neighbour Cartesian velocity field ($\mathbf{u}_{\text{nn}}$) to the local normal-tangential coordinates. This transformed velocity field then underpins the resolvent analysis, enabling focused investigation into boundary layer dynamics and the identification of coherent structures. Throughout this study, we will refer to the transformed velocity field without the subscript $\mathcal{B}$ for the sake of brevity.

    \subsection{Resolvent analysis}\label{sec:RA formulation}

        We use the method of \citealt{herrmann_data-driven_2021} for the resolvent analysis. Briefly, we consider a forced dynamical system represented by a divergence free velocity subspace such that
        \begin{equation}
            \dot{\mathbf{u}} = \mathcal{L}\mathbf{u} + \mathbf{f} \text{ ,}
            \label{eq:dynamical system}
        \end{equation}
        where $\mathbf{u}$ is the velocity field, $\mathcal{L}$ is a dynamical operator driving the evolution of the system, and $\mathbf{f}$ is the forcing term. To determine the linear operator, we consider the evolution of the velocity field in discrete time such that
        \begin{equation}
             \mathbf{U}_{k+1} = A\mathbf{U}_k \text{ ,}
        \end{equation}
        where $ \mathbf{U}_k $ is the state at time $ t_k = k\Delta t $. For convenience, we use $\Delta t = 0.005$, which is the flow field sampling frequency--this rescales time to normalise the definition in section \ref{sec:numerical method} by the flapping frequency. To approximate the operator $ A $, we implement the standard DMD algorithm from \citet{schmid_dynamic_2010} using the open source software package \citet{demo_pydmd_2018}. This algorithm is sufficient for our dataset and is able to handle the number of snapshots we have. We construct flat overlapping matrices $X$ and $Y$ such that
        \begin{align}
            X &= \begin{bmatrix}u_0 & u_1 & \cdots & u_{m-1} & v_0 & v_1 & \cdots & v_{m-1}\end{bmatrix}\\
            Y &= \begin{bmatrix}u_1 & u_2 & \cdots & u_{m} & v_1 & v_2 & \cdots & v_{m}\end{bmatrix} \text{ ,}
        \end{align}
        and employ a dimensionally efficient projection using the pseudoinverse to approximate the linear map. We take the truncated SVD of the matrix $ X $, producing $ X = U\Sigma V^* $ and estimate the linear map such that $A = U^*YV\Sigma^{-1}$. For this study, we truncate the SVD at 6 as we find that this is the largest value that does not introduce decaying modes into the bases. The DMD modes are then computed by taking the eigendecomposition of $A$, such that $ AV = V\Gamma $ and $ A^+W = W\Gamma^* $, where $ A^+ $ is the adjoint of $ A $.

        For the resolvent analysis, we consider equation \ref{eq:dynamical system} in terms of eigenvector coordinates such that $ x(t) = V a(t) $ and $ f(t) = V b(t) $, with $ a, b $ being expansion coefficients. In continuous time, $\Lambda = \ln(\Gamma)/\Delta t$, so the equation \ref{eq:dynamical system} can be written as
        \begin{equation}
            \dot{a} = \Lambda a + b \text{ .}
        \end{equation}
        To maintain the 2-norm's physical meaning, we adjust the inner product via a weighting matrix defined from the Cholesky factorisation of $ V^*V = \widetilde{F}^*\widetilde{F} $. The resolvent operator is then defined as
        \begin{equation}
            \widetilde{F}(-i\omega I - \Lambda)^{-1}\widetilde{F}^{-1} = \Psi_{\widetilde{F}} \Sigma \Phi^{*}_{\widetilde{F}} \text{ ,}
            \label{eq: resolvent operator}
        \end{equation}
        where we have decomposed the resolvent operator into the singular value decomposition, $\Sigma$, and the forcing and response modes, $\Phi^{*}_{\widetilde{F}}$ and $\Psi_{\widetilde{F}}$, respectively. By synthesising resolvent modes into physical coordinates, $ \Phi = V \widetilde{F}^{-1}\Phi_{\widetilde{F}}$ and $ \Psi = V \widetilde{F}^{-1} \Psi_{\widetilde{F}} $ represent the forcing and response modes, respectively.

\section{Results} 
    \subsection{Maximum gain}
        \begin{figure}
            \centering
            \includegraphics{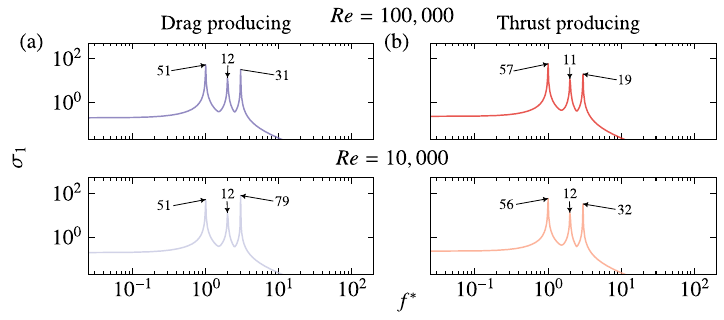}
            \caption{The gain spectra of the four test cases. The upper row is the spectra for the $\Rey=100,000$ case and the bottom row is the spectra for $\Rey=10,000$. The two columns represent the propulsion conditions; (a) drag producing and (b) thrust producing. The annotations indicate the peak values for each of the cases.}
            \label{fig:gain}
        \end{figure}

        
        The dominant time scales of the system dynamics are identified by the forcing frequency resulting in the maximum gain of the resolvent operator. Figure \ref{fig:gain} shows the gain spectra of the resolvent operator for the four test cases where the top and bottom rows differentiate between $\Rey=10,000$ and $100,000$, and the two columns are the propulsion conditions; (a) drag producing and (b) thrust producing. For ease of comparison, we have labelled the values of the local maxima. The two peaks at the first and second harmonics remain largely constant across all the cases, although there is minimal variation between the thrust and the drag cases, with the thrust exhibiting slightly more sensitivity at the first harmonic. The differences in stability between the cases are present in the third harmonic; there is also a significant difference between the thrust and the drag cases in $\Rey$ values. The dynamics at $f^*=3$ are less sensitive to forcing at $\Rey=100,000$ than at $\Rey=10,000$.
        The amplification of this mechanism is greater than in the second harmonic and, for the $\Rey=10,000$ drag case, more than in the first harmonic.

    \subsection{Resolvent modes}\label{sec:resolvent modes}
        \begin{figure}
            \centering
            \includegraphics[width=\textwidth]{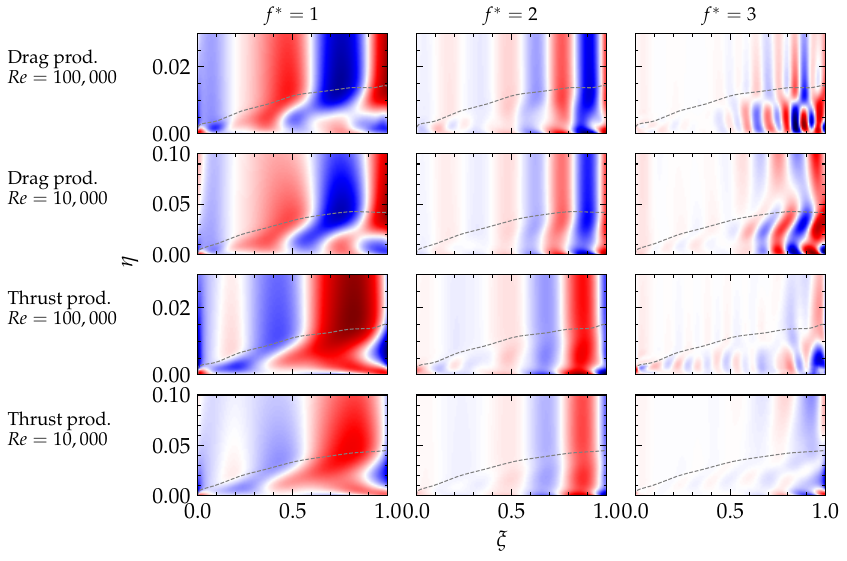}
            \caption{The response modes represented by the normalised tangential velocity ($u_{\xi}/||u_{\xi}||_{\infty}$), for the first three harmonics of the resolvent operator. Each row represents the four distinct test cases and is labelled as appropriate. The colour palette ranges from -1 (blue) to 1 (red). The grey dashed line depicts the boundary-layer thickness.}
            \label{fig:FR modes}
        \end{figure}

        The response modes of the resolvent operator are the mechanisms driving the system, as depicted in figure \ref{fig:FR modes}. These modes correspond to the peaks shown in figure \ref{fig:gain}, representing the first three harmonics of the resolvent operator. The first harmonic, with the exception of the $\Rey=100,000$ drag production case, is the most energetic mode. It exhibits coherent structures in both the thrust and drag regimes, reinforcing the idea that kinematics are the dominant factor. The drag cases differ from the thrust cases by the presence of large fluctuations ($k_\xi \approx 3$) associated with flow reversal in the back half of the foil, which disrupt the propulsive wave. In the thrust-producing cases, these fluctuations are absent, and the energy amplification across the back two-thirds of the foil is more uniform. The time averaged boundary-layer thickness is indicated by the grey dashed line in figure \ref{fig:FR modes}. This was identified via the shear-threshold method \citep{uzun2021simulation}, using a threshold of 0.001. The unique thrust- and drag-producing structures in the first harmonic scale with the boundary-layer thickness across the viscous regimes implying a geometric self-similarity of the energetic structures.

        The second harmonic predominantly exhibits waves that do not depend on $\eta$ or the boundary-layer thickness and increase in amplitude towards the tail, matching the increasing motion amplitude envelope along the body length. The streamwise wavenumber of these structures is twice the prescribed locomotive wave for each case; for the drag-producing cases ($k_\xi \approx 3.5$) and for the thrust-producing cases ($k_\xi \approx 2.5$). At the nose of the foil, the $\Rey=100,000$ cases display fluctuations associated with shedding that are absent in the $\Rey=10,000$ cases.
        Finally, the third harmonic shows positive and negative fluctuations that intensify towards the tail in both regimes, though they are more pronounced in the drag regime. The amplitude of these fluctuations is significantly higher within the boundary-layer region. This implies that the third harmonic is less correlated with body motion and is driven by boundary-layer dynamics. Furthermore, the wavenumber of the structures does not seem to correlate with any integer multiple of the propulsive wave.
        Given the high amplification of the third harmonic, its modification offers an avenue to control that does not depend on changing the kinematics.
    
    \subsection{Roughness comparison}

        \begin{figure}
            \centering
            \includegraphics[width=\textwidth]{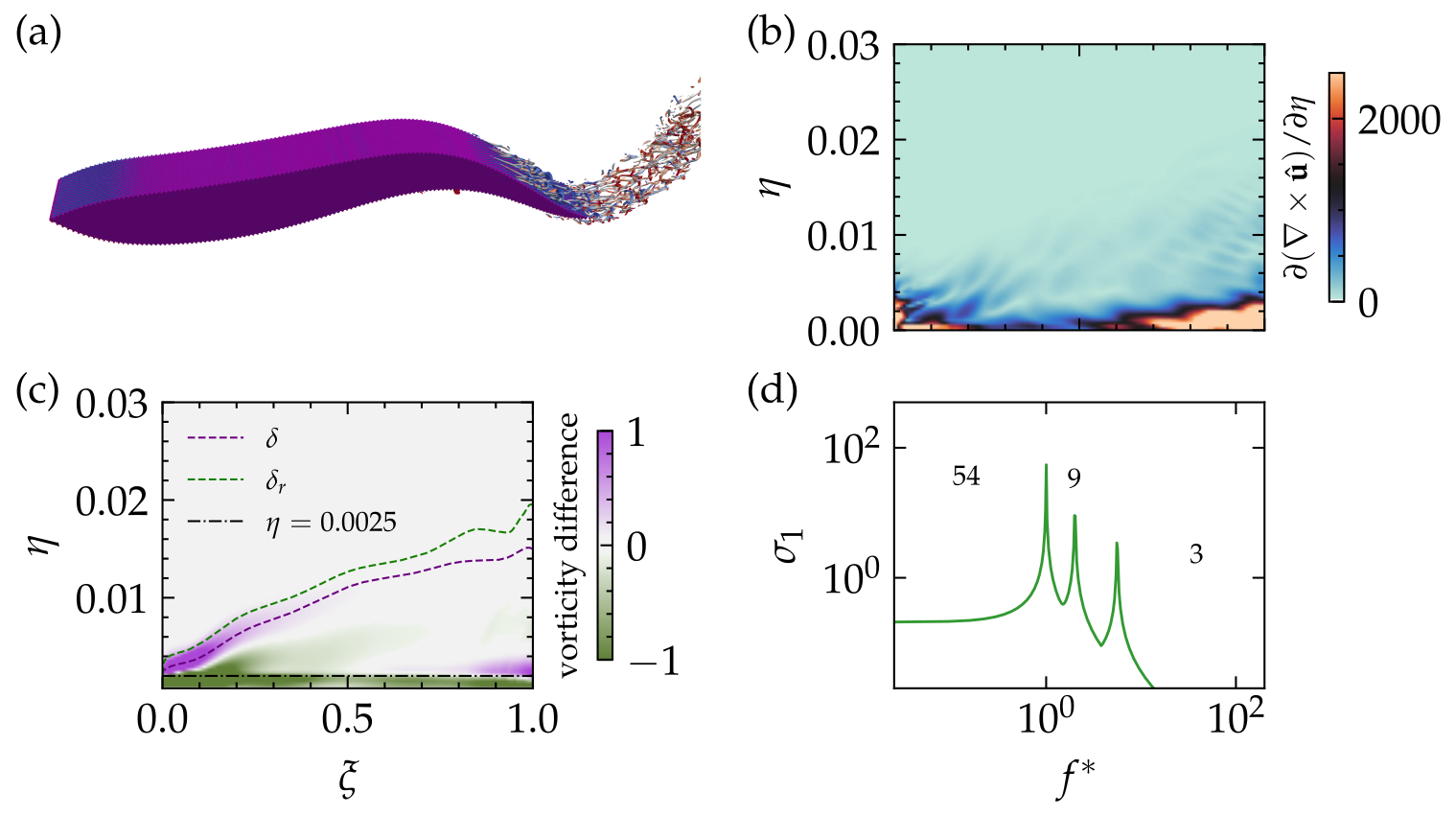}
            \caption{Study on the effect roughness has on the third harmonic of the $\Rey=100,000$ thrust-producing case. (a) is a Q-criterion contour of $10^{-4}$ coloured by $\omega_z$ for the flow over the thin foil with egg-carton roughness. (b) shows the absolute value of the normalised vorticity flux for the smooth foil forcing mode at the third harmonic. (c) is the difference in vorticity flux between the smooth and the rough foil with the boundary-layer thicknesses and a black, dash-dot line marking the height of the bumps. (d) is the gain spectra for the rough foil resolvent operator.}
            \label{fig:control}
        \end{figure}

        As a final example, we demonstrate the ability of the new dynamic resolvent analysis to study the impact of roughness on the amplification of the dominant mechanisms in the flow. Specifically, we look at the change in amplification of the mechanism at the third harmonic which we have identified as one decorrelated to the kinematics (section \ref{sec:resolvent modes}). We make the link to the vorticity flux as the potential control mechanism because vorticity flux-based control has been shown to reduce turbulent fluctuations in a turbulent channel by $40\%$ \citep{koumoutsakos_vorticity_1999}.
        
        In figure \ref{fig:control}b, we show the absolute value of the vorticity flux $\partial (\mathbf{\nabla}\times \mathbf{u})/\partial \eta$ in the forcing mode at the third harmonic for the $\Rey=100,000$ thrust-producing case. The forcing mode consists of significant areas of high vorticity flux near the wall.
        We add an egg-carton roughness by adding a surface excursion such that
        \begin{equation}
            y(x,z)= \begin{cases}
            h \sin\Big(\frac{2\pi x}{\lambda}\Big)\cos\Big(\frac{2\pi z}{\lambda}\Big), &  \text{for } y > 0 \\
            h \sin\Big(\frac{2\pi x}{\lambda}-\pi\Big)\cos\Big(\frac{2\pi z}{\lambda}\Big), & \text{for  } y < 0
            \end{cases}
            \label{eq:roughness}
        \end{equation}
        where h is the bump amplitude, set at $0.002$, and $\lambda$ is the bump wavelength, set at $1/128$ to ensure the roughness is small but still resolved by the grid. We include 16 roughness wavelengths across the span of the foil. Figure \ref{fig:control}a shows the flow of the rough foil, with the Q-criterion coloured by the spanwise vorticity. The flow remains similar to the smooth foil, although the structures are offset by the roughness amplitude and break down before being ejected from the tail.
        
        The vorticity flux in the roughness sublayer is greatly reduced by the roughness: Figure \ref{fig:control} c shows the difference between rough and smooth, and the region within the roughness height is marked with a black dashed dotted line. The reduction in flux directly opposes the mechanisms present in the forcing mode of the smooth foil (figure \ref{fig:control}a).
        In figure \ref{fig:control}c, we also mark the boundary-layer thicknesses for the smooth ($\delta$) and the rough ($\delta_r$) foil which is shifted in the rough case. Near the nose, there is a region where the vorticity flux is greatly increased; this is a side effect of the control method.
        The reduction in vorticity flux corresponds to a significant reduction in the amplification of the mechanism at the third harmonic, as shown by the gain spectra in figure \ref{fig:control}d. The amplification of the first and second harmonics is also slightly reduced. The comparison shows that roughness can be used to control the amplification of key mechanisms in the flow.


\section{Conclusions}

    In this study, we applied data-driven resolvent analysis to explore the boundary-layer dynamics of a foil undergoing undulatory propulsion, across thrust- and drag-producing regimes, and at realistic Reynolds numbers. Using a novel coordinate transformation, we were able to perform the data-driven resolvent analysis technique.
    The analysis revealed that the kinematics of the system are the dominant factor driving the physics. The thrust- and drag-producing cases exhibit distinct structures that scale with the boundary-layer thickness. The drag producing cases show large fluctuations associated with flow reversal in the back half of the foil, disrupting the propulsive wave.
    Finally, we provide a comparison to a foil with egg-carton roughness. The roughness reduces the amplification of the mechanism at the third harmonic, which is less correlated with body motion. This demonstrates the potential for roughness to control the amplification of key mechanisms in the flow.
    By extending resolvent analysis to deforming bodies with non-zero thickness, this work opens new avenues for future research in understanding propulsion mechanisms.
    
\bibliography{references}
\bibliographystyle{jfm}

\end{document}